\documentclass[fleqn,usenatbib]{mnras}
\usepackage{newtxtext,newtxmath}
\usepackage[T1]{fontenc}
\usepackage{ae,aecompl}
\usepackage{graphicx}
\usepackage{amsmath}	
\usepackage{amssymb}	
\usepackage{xcolor}
\usepackage[symbol]{footmisc}

\title[CDM and SIDM simulations of Antlia 2]{Simulating the ``hidden giant" in cold and self-interacting dark matter models}

\author[Sameie et al.]{Omid Sameie,$^{1,3}$ \thanks{E-mail: sameie@utexas.edu}
Sukanya Chakrabarti,$^{2}$
Hai-Bo Yu,$^{3}$
Michael Boylan-Kolchin,$^{1}$
\newauthor Mark Vogelsberger,$^{4}$
Jes\'us Zavala,$^{5}$
and Lars Hernquist$^{6}$
\\
$^{1}$Department of Astronomy, The University of Texas Austin, 2515 Speedway, Stop C1400, Austin, TX 78712 USA\\
$^{2}$School of Physics and Astronomy, Rochester Institute of Technology, Rochester, NY 14623 USA\\
$^{3}$Department of Physics and Astronomy, University of California, Riverside, CA 92521 USA\\
$^{4}$Department of Physics, Kavli Institute for Astrophysics and Space Research, Massachusetts Institute of Technology,\\ Cambridge, MA 02139, USA\\
$^{5}$Center for Astrophysics and Cosmology, Science Institute, University of Iceland, Dunhagi 5, 107 Reykjavik\\
$^{6}$Harvard-Smithsonian Center for Astrophysics, 60 Garden Street, Cambridge, MA 02138 USA
}

\date{Accepted XXX. Received YYY; in original form ZZZ}

\pubyear{2020}

\begin{document}
\label{firstpage}
\pagerange{\pageref{firstpage}--\pageref{lastpage}}
\maketitle

\begin{abstract}
We perform a series of controlled N-body simulations to study realizations of the recently discovered Antlia 2 galaxy in cold dark matter (CDM) and self-interacting dark matter (SIDM) scenarios. Our simulations contain six benchmark models, where we vary the initial halo concentration and the self-scattering cross section. We adopt well-motivated initial stellar and halo masses, and our fiducial orbit has a small pericenter. After evolving in the Milky Way's tidal field, the simulated galaxies experience significant mass loss and their stellar distributions expand accordingly. These tidal effects are more prominent if the initial halo concentration is lower and if the self-scattering cross section is larger. Our results show that Antlia 2-like galaxies could be realized in CDM if the halo concentration is low and the stellar distribution is diffuse at the infall time, while these conditions could be relaxed in SIDM. We also find all the simulated galaxies predict approximately the same stellar velocity dispersion after imposing selection criteria for stellar particles. This has important implications for testing dark matter models using tidally disturbed systems.

\end{abstract}


\section{Introduction}
\label{sec:intro}

The newly discovered Antlia 2 satellite galaxy is the lowest surface brightness galaxy known to date \citep{torrealba2019}.  Its half-light radius is $\sim2.9~{\rm kpc}$, making it two orders of magnitude more diffuse than ultra-diffuse galaxies \citep{Koda:2015gwa}. It is located near the Galactic plane, at a galactic latitude of $\sim 11^{\circ}$. Because of these unique properties, the Antlia 2 galaxy has been dubbed ``hidden giant." In fact, \citet{bullock2010} predicted the existence of this type of so-called stealth galaxies, which have similar luminosities but more extended stellar distributions, compared to ultra-faint dwarf galaxies.

\cite{chakrabarti2019} used Gaia proper motions \citep{gaia2018} to derive orbital parameters of the Antlia 2 galaxy. They also calculated its dynamical effect on the outer HI disk of the Milky Way and found orbits with low pericenters ($\sim 10$ kpc) could match the observed disturbances in the HI disk \citep{levine2006}. Interestingly, \cite{chakrabarti2009,chakrabarti2011} proposed that the HI disturbances could be explained by the existence of a subhalo as a perturber, and its predicted radial location at the present time is consistent with that of Antlia 2.

\citet{torrealba2019} used controlled N-body simulations to explore the origin of Antlia 2 in the Milky Way's tidal field, and  they considered both cuspy and cored dark matter density profiles. To match the observations, they found that the cuspy halo requires unreasonable model parameters. For example, the adopted initial halo mass is $\sim5\times 10^8\, \text{M}_\odot$, which is too low to form galaxies \citep{okamoto2008,benitez2020}, and the ratio of stellar-to-halo scale radii at the infall is $\sim6$, too large to be produced in hydrodynamical simulations even with strong baryonic feedback \citep{fitts2017,wheeler2019,lazar2020}. For their cored profile, the initial halo mass is much larger, $\sim10^{10}~{\rm M_\odot}$, and the assumed radius of the stellar distribution is comparable to the halo's scale radius. They found that Antlia 2 favors a cored halo than a cuspy one. 

The indication that Antlia 2 may prefer a cored halo profile is of significant interest in the context of the core vs. cusp issue of cold dark matter (CDM), i.e., many dwarf spiral galaxies favor a dark matter density core over a cusp as predicted in CDM-only simulations \citep{Dubinski:1991bm,navarro1997}; see~\cite{deBlok:2009sp,bullock2017,tulin2018} for reviews. Recent simulations show that baryonic feedback could produce density cores \citep{Read:2004xc,Governato:2009bg,chan2015,Santos-Santos2018,fitts2019} and the dark matter core size could be as large as $\sim3~{\rm kpc}$ for a $10^{10}~{\rm M_\odot}$ CDM halo \citep{lazar2020}. On the other hand, the cores could also form if dark matter has strong self-interactions~\citep{spergel2000,feng2009,arkani2009,loeb2011,tulin2013,Kaplinghat:2015aga}. It has been shown that this self-interacting dark matter (SIDM) scenario could explain diverse dark matter distributions in the field galaxies \citep{creasey2017,kamada2017,ren2019}, dwarf galaxies of the Milky Way \citep{vogel2012,zavala2013,Valli:2017ktb,kahlhoefer2019,sameie2020} and ultra diffuse galaxies \citep{Yang:2020iya}. It is interesting to see whether Antlia 2 could shed further light on the nature of dark matter. 

In this work, we study realizations of Antlia 2 using controlled N-body simulations in both CDM and SIDM scenarios. The simulated satellites contain halo and stellar components and their masses are well motivated by the stellar-halo mass \citep{moster2013,behroozi2013} and mass-metallicity \citep{kirby2013} relations. For the CDM simulations, we vary initial halo concentration, while for the SIDM simulations we fix the concentration, but vary the dark matter self-scattering cross section. Our orbital distribution is derived from the observed Gaia proper motions of Antlia 2, and our fiducial orbit has a low pericenter and is nearly co-planar, like the orbit that \citet{chakrabarti2019} found matched the disturbances in the outer HI disk of the Milky Way. We further compare our simulations to Antlia 2's observables reported in \cite{torrealba2019}, including the observed line-of-sight (LoS) velocity dispersion, half-light radius, stellar mass, as well as dynamical mass within the half-light radius. 

As we will show, tidal interactions play an important role in shaping dark matter and stellar distributions of our simulated galaxies in both CDM and SIDM. In contrast to the early results \citep{torrealba2019}, we find Antlia 2-like galaxies could be realized in CDM if the initial halo has a low concentration and the initial stellar distribution is highly diffuse, while in SIDM a higher halo concentration and a compact stellar distribution are also allowed. All our simulated galaxies have approximately the same LoS stellar velocity dispersion after a realistic velocity cut, although their inner dark matter densities are different. We discuss its implications for discriminating between CDM and SIDM models. We also investigate gravitational boundedness on the computation of the stellar kinematics and the impact of different orbital trajectories. 

The paper is organized as follows.  In \S \ref{sec:sims}, we discuss our simulation setup. In \S \ref{sec:results}, we present our results and discuss their implications. We conclude in \S \ref{sec:conclusion}.

\section{Simulations}\label{sec:simulations}
\label{sec:sims} 

We perform our simulations using a modified version of the code {\sc Arepo} \citep{springel2010}, which includes a module implementing the dark matter self-interactions \citep{vogel2012,vogel2016,vogelsberger2019}. Following \cite{chakrabarti2019}, we model the host including both halo and stellar components with a Hernquist profile \citep{hernquist1990},
\begin{equation}
    \Phi_{\rm h}\, (r)=-\frac{G M_{\rm h}}{r+r_{\rm h}},
    \label{eq:hernquist}
\end{equation}
where $M_{\rm h}=1.86\times10^{12}\, {\rm M}_\odot$ and $r_{\rm h}=36.1~{\rm kpc}$ are total mass and characteristic length scale, respectively. The corresponding virial mass of the halo is $M_{\rm v}\approx1.4\times10^{12}\, {\rm M}_\odot$. This is comparable to recent mass models of the Milky Way inferred from observations \citep{mbk2013,fritz2018,watkins2019,deason2019,posti2019}. In our simulations, we assume the host potential given in Eq.~\ref{eq:hernquist} is static.

We use a Navarro-Frenk-White profile \citep{navarro1997} to model the initial dark matter distribution of the simulated galaxies,
\begin{equation}
    \rho_{\rm dm}\, (r)=\frac{\rho_{\rm s}}{ r/r_{\rm s} \big(1+r/r_{\rm s}\big)^2 },
\end{equation}
where $\rho_{\rm s}$ and $r_{\rm s}$ are the characteristic density and length scale. Equivalently, we can use the virial mass $M_{\rm v}$ and halo concentration $c_{\rm v}\equiv r_{\rm v}/r_{\rm s}$ to specify a halo, where $r_{\rm v}$ is the halo's virial radius. We consider six benchmark models for our simulations. We fix $M_{\rm v}=10^{10}~{\rm M_\odot}$, estimated from the stellar-halo mass relation \citep{moster2013,behroozi2013}. For CDM, we choose $c_{\rm v}=13,\, 10$ and $8$, corresponding to the median of $c_{\rm v}(M_{\rm v})$ relationship and $1\sigma$ and $2\sigma$ deviations below the median concentration at fixed mass found in cosmological simulations at $z=0$ \citep{dutton2014}. Compared to \cite{torrealba2019}, where $c_{\rm v}=15.9$ is fixed, we explore a broader range of halo models. For our SIDM simulations, we set $c_{\rm v}=10$, but vary the self-scattering cross section as $\sigma/m=1,\; 3 $ and $5\, {\rm cm^2/g}$ (hereafter denoted as SIDM1, SIDM3 and SIDM5, respectively). These cross sections are motivated to address CDM's small-scale issues and pass observational tests on galactic scales \citep{tulin2018}.

The initial stellar component of the satellites is modeled with a Plummer profile \citep{plummer1911},
\begin{equation}
    \rho_{\rm p}\, (r)=\bigg(\frac{3M_{\rm p}}{4\pi}\bigg) \frac{r_{\rm p}^2}{\big(r_{\rm p}^2+r^2\big)^{\frac{5}{2}}},
\end{equation}
where $M_{\rm p}$ is the mass and $r_{\rm p}$ is the scale length of the stellar distribution. In order to accommodate stellar mass loss after tidal evolution, we assume $M_{\rm p}=3\times 10^7~{\rm M_\odot}$, which is on the higher limit of the mass-metallicity relation \citep{kirby2013}. We choose the length scale as $r_{\rm p}=2~{\rm kpc}$. This is motivated by recent hydrodynamical simulations in \citet{lazar2020}, where they showed a 3D half-mass radius of $r_{1/2}\sim3~{\rm kpc}$ could be achieved for a $10^{10}~{\rm M_\odot}$ halo. Note for a Plummer profile the 2D projected half-mass radius is $R_{1/2}=r_{\rm p}$, and $r_{1/2}\approx 4/3R_{1/2}$ \citep{wolf2010}.

The simulated galaxies consist of $5$ million dark matter and $1$ million star particles. The mass resolutions are $3.6\times 10^3\, {\rm M}_\odot$ and $30\, {\rm M}_\odot$ for the dark matter and stars. The Plummer-equivalent softening length is $\epsilon_{\rm p}=25\, \text{pc}$. The initial conditions for the simulated galaxies are generated using the publicly available code {\sc SPHERIC} \citep{gk2013}.

We follow \citet{chakrabarti2019} and obtain the orbits for the simulated galaxies. The initial position and velocities are derived from backward integrating the observed Gaia proper motions of the Antlia 2 dwarf for $8~{\rm Gyr}$. We further sample the errors in the proper motions to derive an orbital distribution. The orbit we consider has apocenter and pericenter radii of $\sim 200~{\rm kpc}$ and $13~{\rm  kpc}$. 

\begin{figure}
    \centering
    \includegraphics[width=\columnwidth]{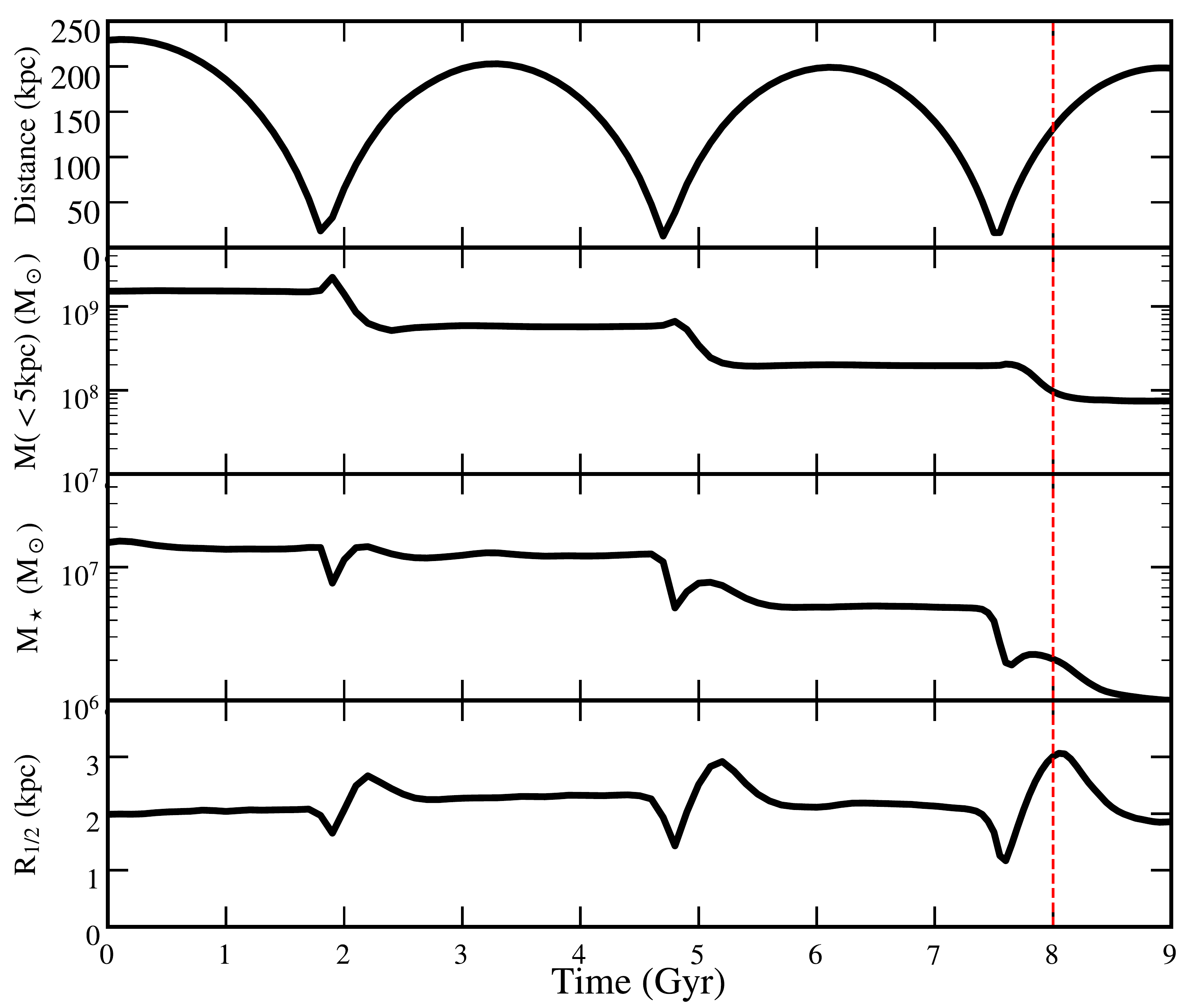}
    \caption{Galaxy properties vs. time for the CDM satellite with the initial concentration of $c_{\rm v}=10$. From top to bottom panels: the distance from the host, dynamical mass within inner $5~{\rm kpc}$, stellar mass within inner $7~{\rm kpc}$, and the stellar half-mass radius. The vertical dashed red line denotes $t=8\, {\rm Gyr}$, at which we evaluate observables of our simulated satellites and compare them with observations of Antlia 2.}
    \label{fig:evolution}
\end{figure}

\section{Results}
\label{sec:results}

\subsection{Tidal evolution and dark matter distributions}

We investigate the tidal evolution of the $c_{\rm v}=10$ CDM galaxy. Fig. \ref{fig:evolution} shows the characteristic observables vs. time, including the distance from the host, dynamical mass, stellar mass and 2D half-mass radius. In computing the dynamical mass, we sum both halo and stellar masses within an inner $5~{\rm kpc}$ radius of the galaxy. To estimate the stellar mass and half-mass radius, we first choose stars that pass a velocity cut of $|v_{\rm los}-V_{\rm bulk}|\leq 12$ km/s, where $v_{\rm los}$ is the LoS velocity of the stars and $V_{\rm bulk}$ is the bulk velocity of the satellite; we will discuss velocity cut in detail later. Then, the stellar mass and half-mass radius are evaluated by fitting a Plummer profile to the simulated stellar particles inside of a $7~{\rm kpc}$ radius at each snapshot.

We estimate Antlia 2's infall to be around $8~{\rm Gyr}$ ago, based on orbit-integration calculations \citep{chakrabarti2019}. From Fig.~\ref{fig:evolution}, we see the distance of the simulated galaxy from its host at $t=8~{\rm Gyr}$ is $135~{\rm kpc}$, in good agreement with the measured value of $132~{\rm kpc}$ for Antlia 2 \citep{torrealba2019}. The dynamical mass decreases substantially at each pericenter passage, especially at the first two, due to tidal stripping. After $8~{\rm Gyr}$ of evolution, the satellite has lost more than $90\%$ of its initial mass. In contrast, the tidal mass loss of stars only becomes more significant on the later stages. This is because the tidal stripping is an outside-in process, i.e., it first removes most of the mass in the outskirts dominated by dark matter, and then continues to strip more concentrated stellar mass in the inner regions \citep[see also][]{penarrubia2008,penarrubia2010}. In addition, there is an oscillatory trend in the half-mass radius over the course of the evolution. The stellar distribution progressively becomes more diffuse after each pericenter passage. At $t=8~{\rm Gyr}$, the 2D half-mass radius is maximized, $R_{1/2}\sim3~{\rm kpc}$, which is close to the measured value, and is $50\%$ larger than its initial value.

\begin{figure*}
    \centering
    \includegraphics[width=\columnwidth]{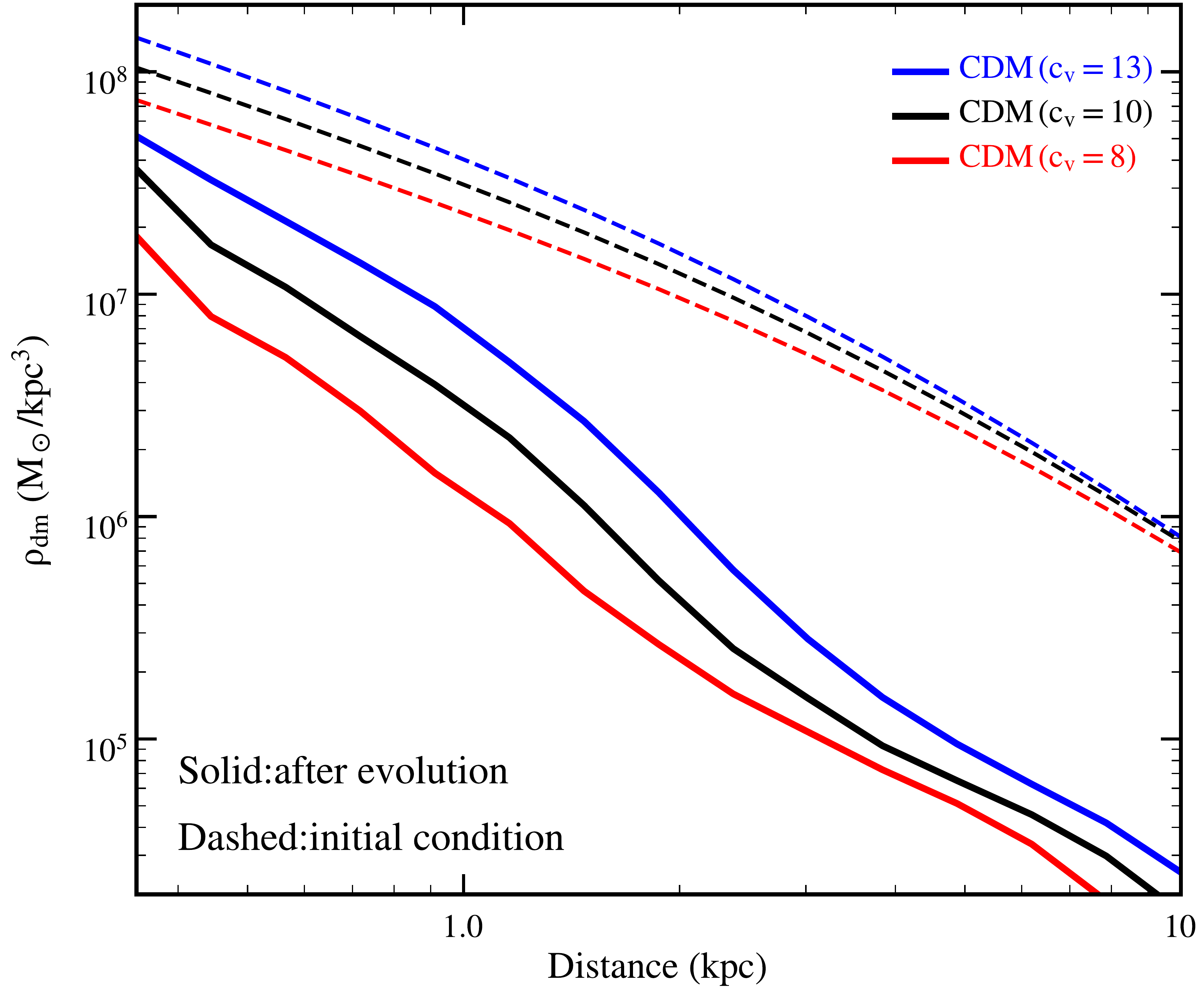}
    \includegraphics[width=\columnwidth]{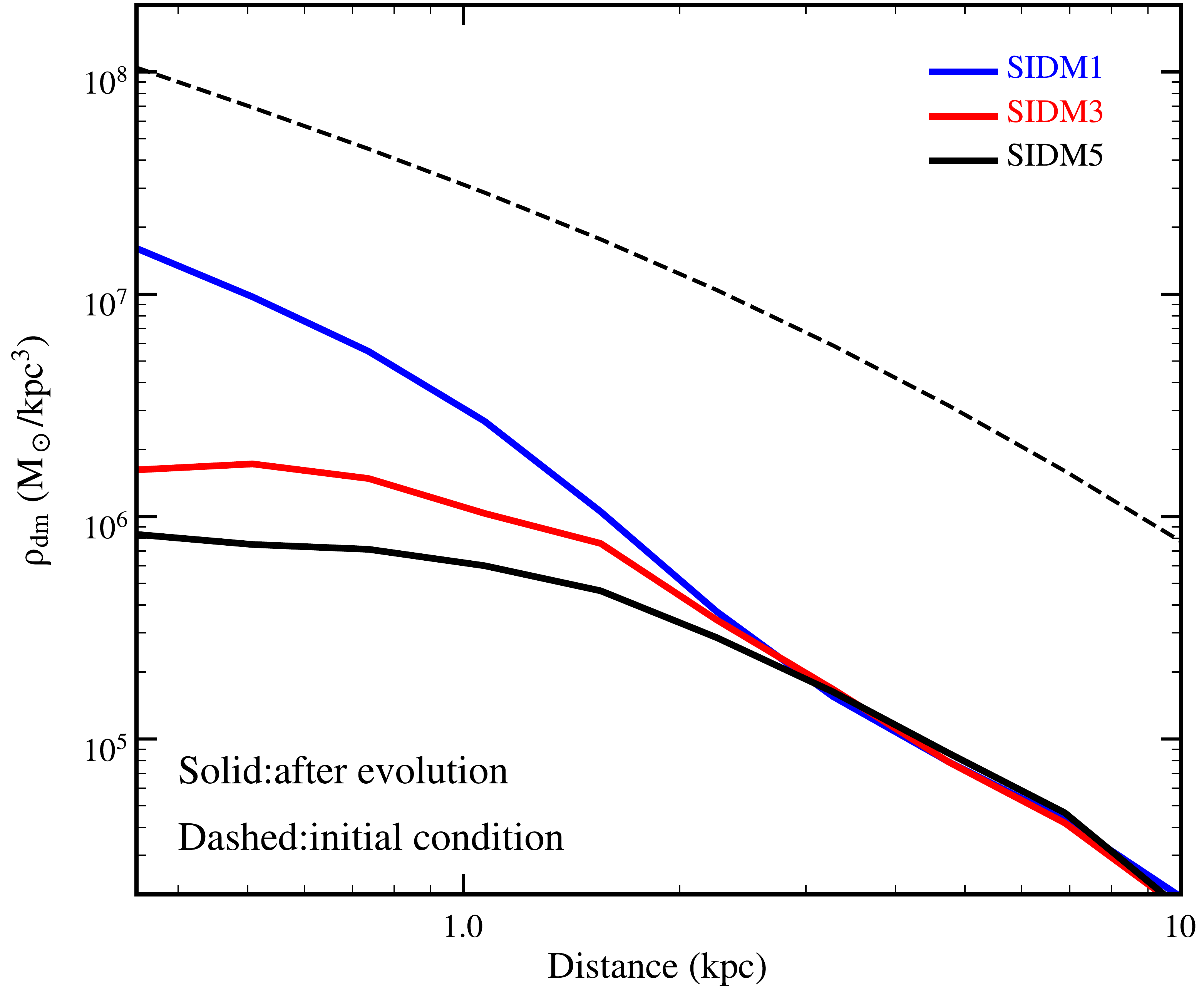}
    \caption{Dark matter density profiles of the simulated satellites for the CDM (left) and SIDM (right) models after $8~{\rm Gyr}$ of evolution in the Milky Way's tidal field (solid), as well as their initial conditions (dashed).}
    \label{fig:density}
\end{figure*}

Fig. \ref{fig:density} shows dark matter density profiles at $t=8~{\rm Gyr}$ for the CDM (left) and SIDM (right) models, together with their initial conditions. All the haloes experience severe mass loss due to the tidal stripping from the host potential and the significance of tidal interactions is more prominent if the concentration is lower or the SIDM cross section is larger. The CDM haloes still retain their inner cuspy structure after tidal evolution \citep{kazantzidis2004, penarrubia2008}. This is true even for the halo with the lowest concentration in our simulations ($c_{\rm v}=8$), which is $2\sigma$ below the median of the $c_{\rm v}(M_{\rm v})$ relation \citep{dutton2014}. For the SIDM haloes, density cores form and the core size increases with the self-scattering cross section \citep{sameie2020}.

\subsection{Line-of-sight velocity dispersion profiles}

\label{sec:LoS-veldisp}

The observed LoS velocity dispersion of Antlia 2 provides an important test for our simulations. A fair comparison requires a good understanding of selection criteria for the stellar members. \citet{torrealba2019} selected $221$ spectroscopic members of Antlia 2 having the best measurements of radial velocities with uncertainties better than $10~{\rm km/s}$ for their kinematic analysis; see their Table 2. All the star members fall inside an on-the-sky distance of $3.2\;{\rm kpc}$. The velocity distribution of the spectroscopic members has a peak at $\sim 290\;{\rm km/s}$, which is the bulk velocity of Antlia 2, and a spread of $20\;{\rm km/s}$. They used a foreground model that excludes star members with radial velocities higher or lower than this limit as foreground contamination.  

\begin{figure*}
    \centering
    \includegraphics[width=\columnwidth]{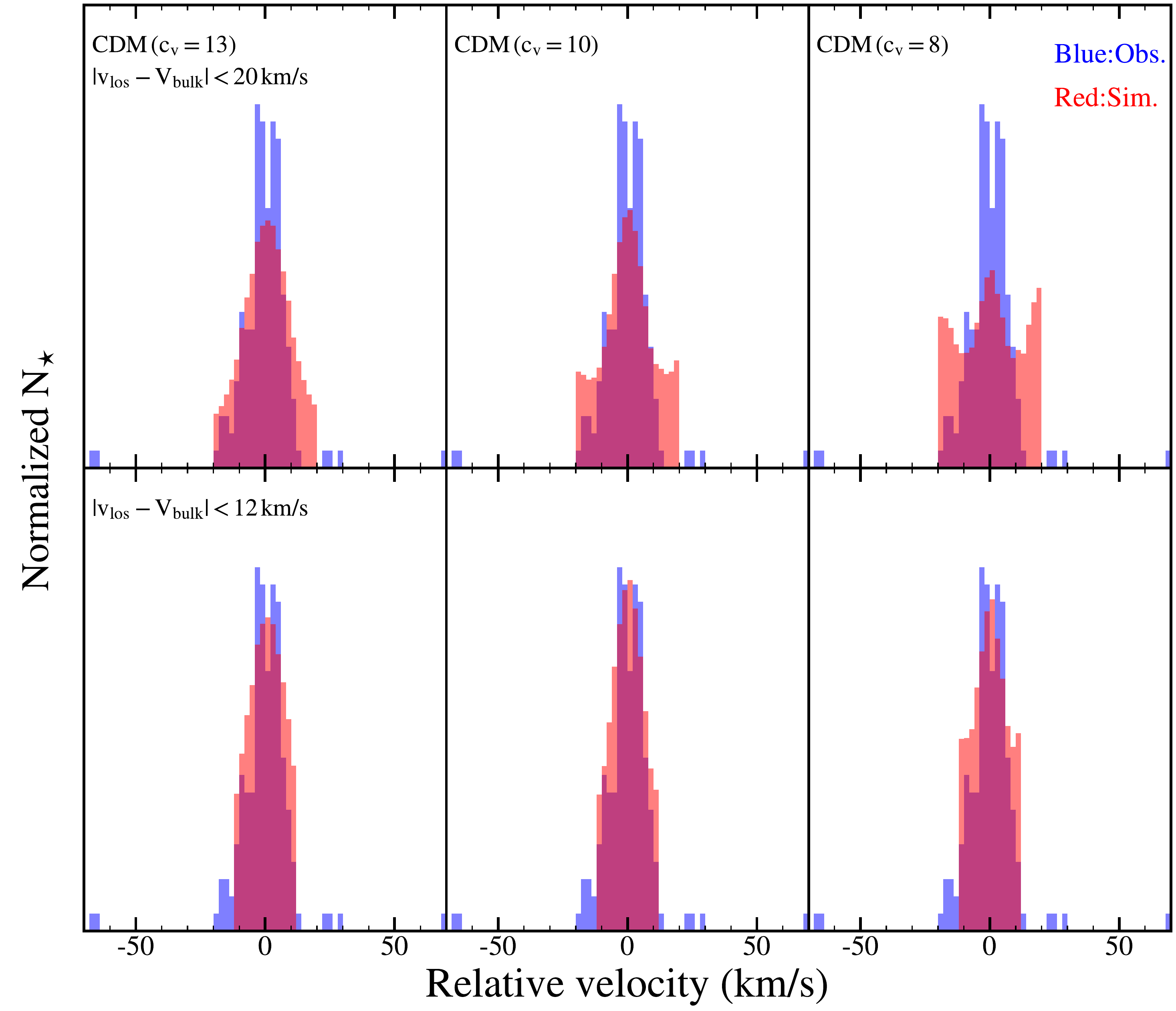}
    \includegraphics[width=\columnwidth]{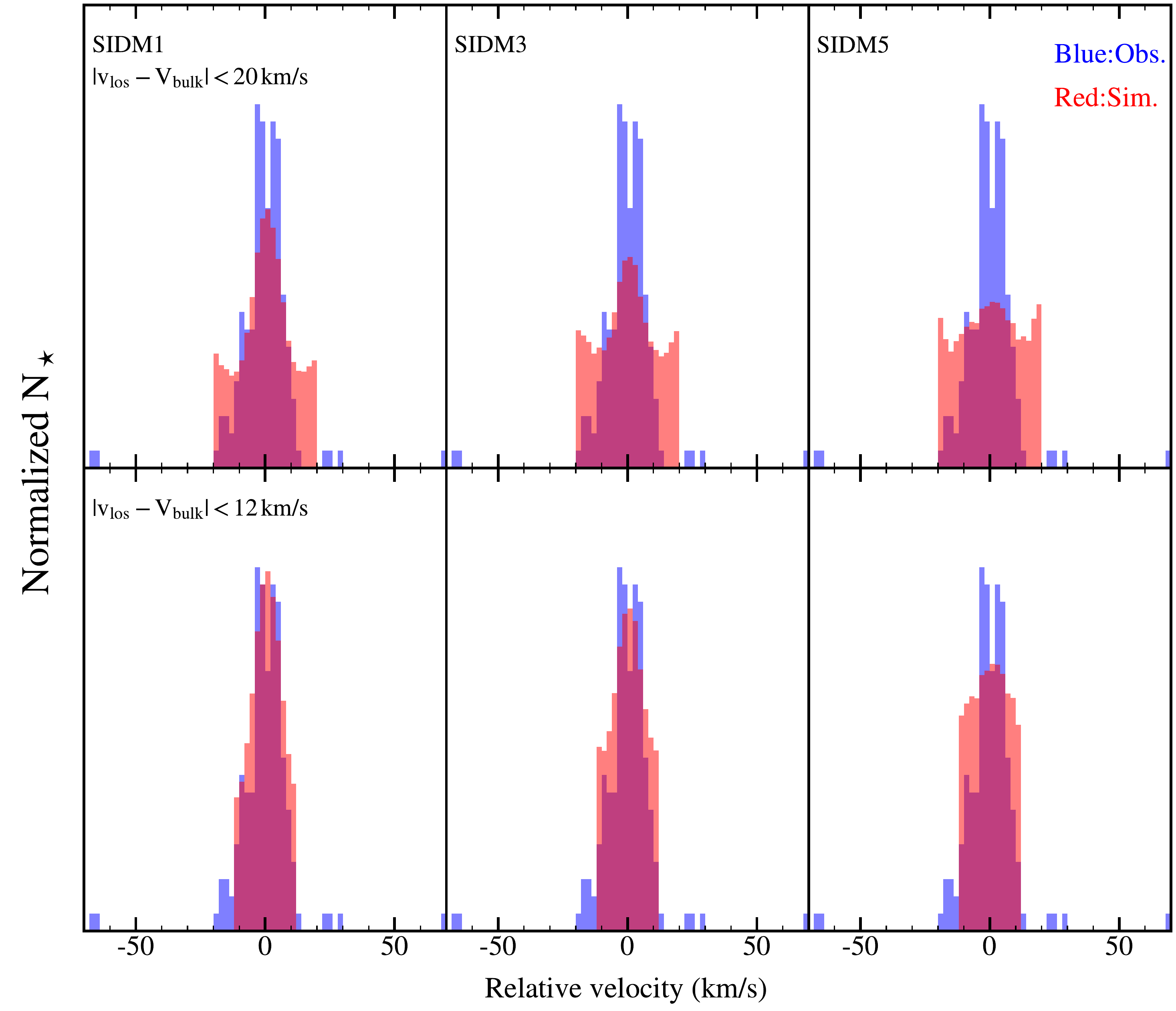}
    \caption{Stellar velocity distributions (red histograms) of the simulated satellites for the CDM (left) and SIDM (right) models with a loose velocity cut (top) and a tight cut (bottom). For comparison, the measured distribution of Antlia 2 is also included  \citep[blue histograms;][]{torrealba2019}.
    }
    \label{fig:histogram}
\end{figure*}

We select simulated stars within a 2D projected distance of $3.2\;{\rm kpc}$ to be consistent with \citet{torrealba2019}. Fig. \ref{fig:histogram} shows distributions of LoS velocities for the CDM and SIDM simulations, after imposing a loose velocity cut $|v_{\rm los}-V_{\rm bulk}|<20~{\rm km/s}$ and a tight cut $|v_{\rm los}-V_{\rm bulk}|<12~{\rm km/s}$, together with the distribution of observed spectroscopic members from \cite{torrealba2019}. With the loose velocity cut, the stellar velocity distributions of the simulated galaxies reveal two side peaks, which are remnants of the tidal debris of stellar particles in our simulations. The peaks are higher for the haloes with lower initial $c_{\rm v}$ or higher $\sigma/m$, as their potentials are shallower and stellar orbits are more likely to be tidally disturbed. Our tight velocity cut removes some of the remnants and the resulting velocity distributions of stars show good agreement with the observations.

\begin{figure*}
    \centering
    \includegraphics[width=\columnwidth]{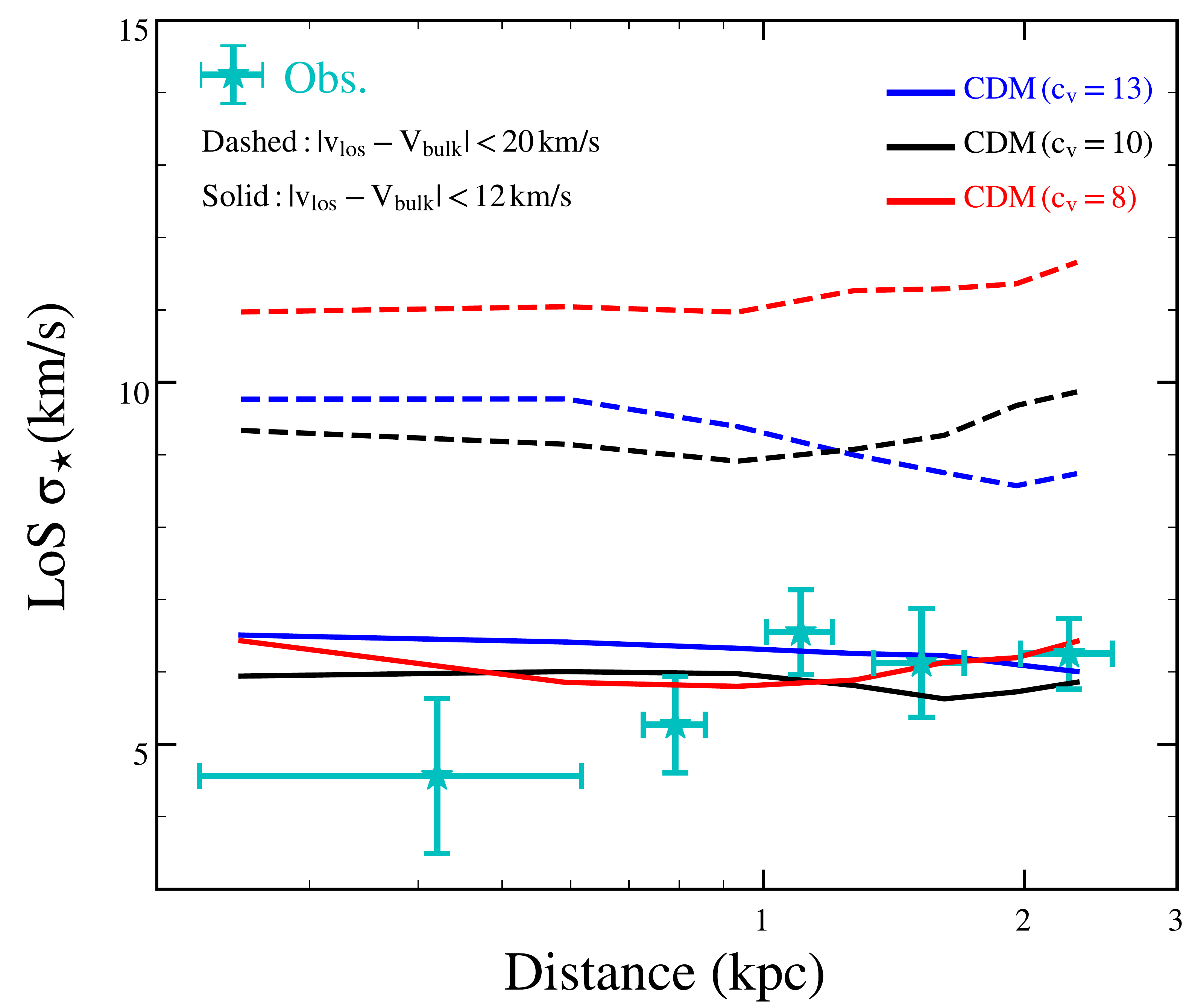}
    \includegraphics[width=\columnwidth]{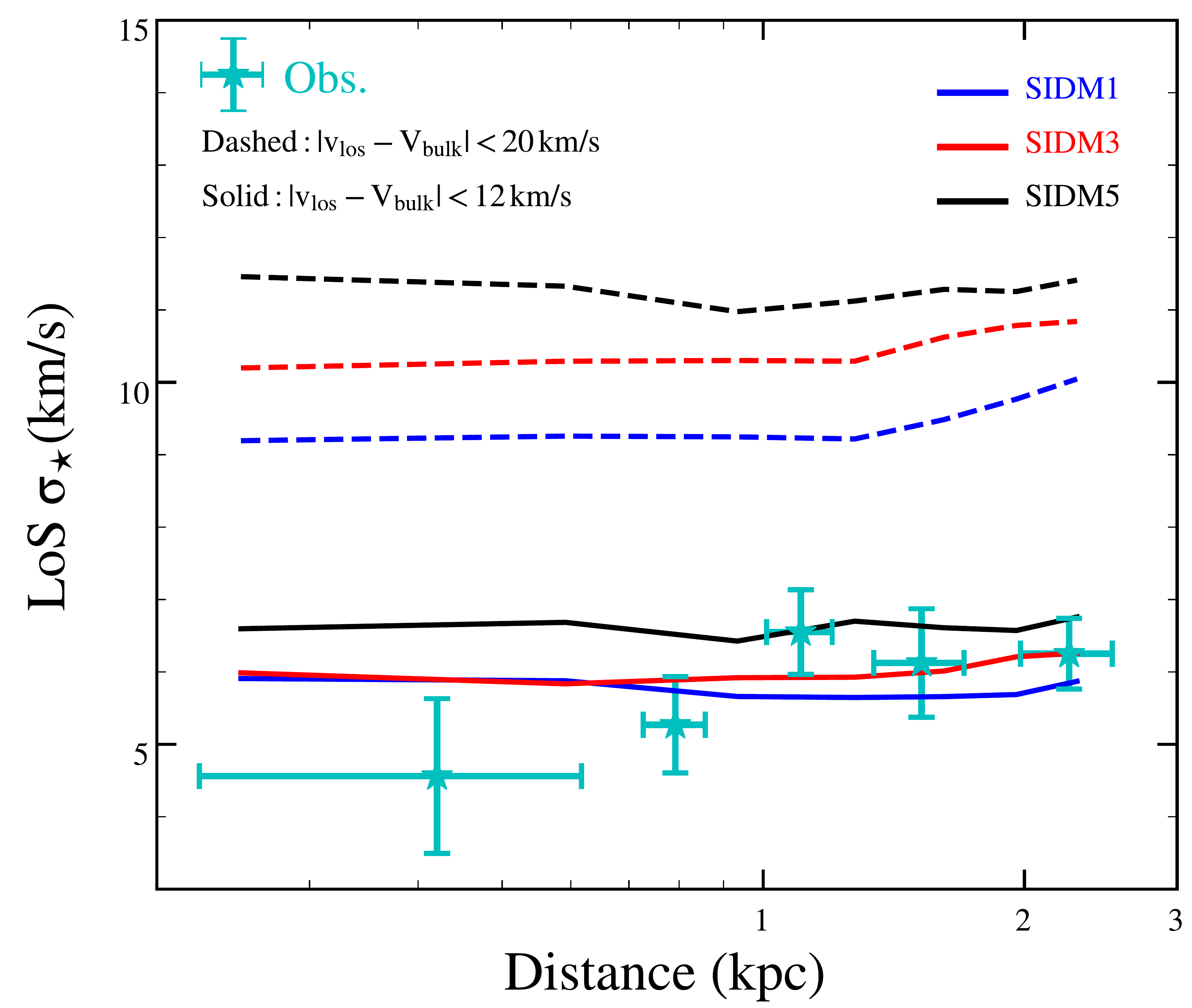}
    \caption{Stellar LoS velocity dispersion for the CDM (left) and SIDM (right) simulations with the loose (dashed) and tight velocity (solid) cuts. The cyan points denote measurements of Antlia 2, which are computed using the data complied in \citet{torrealba2019}}
    \label{fig:sigmaV}
\end{figure*}

\begin{figure*}
    \centering
    \includegraphics[width=\columnwidth]{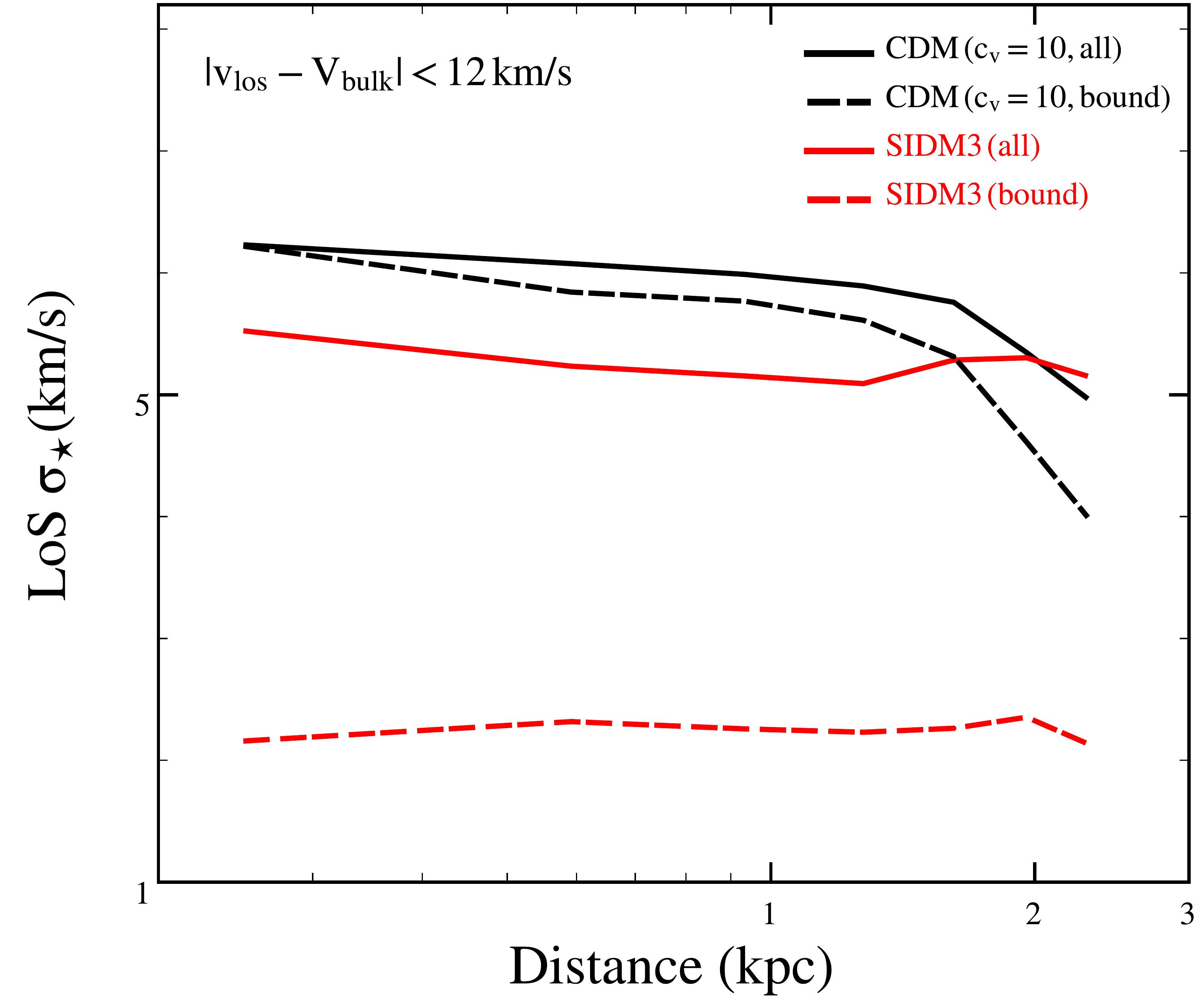}
        \includegraphics[width=\columnwidth]{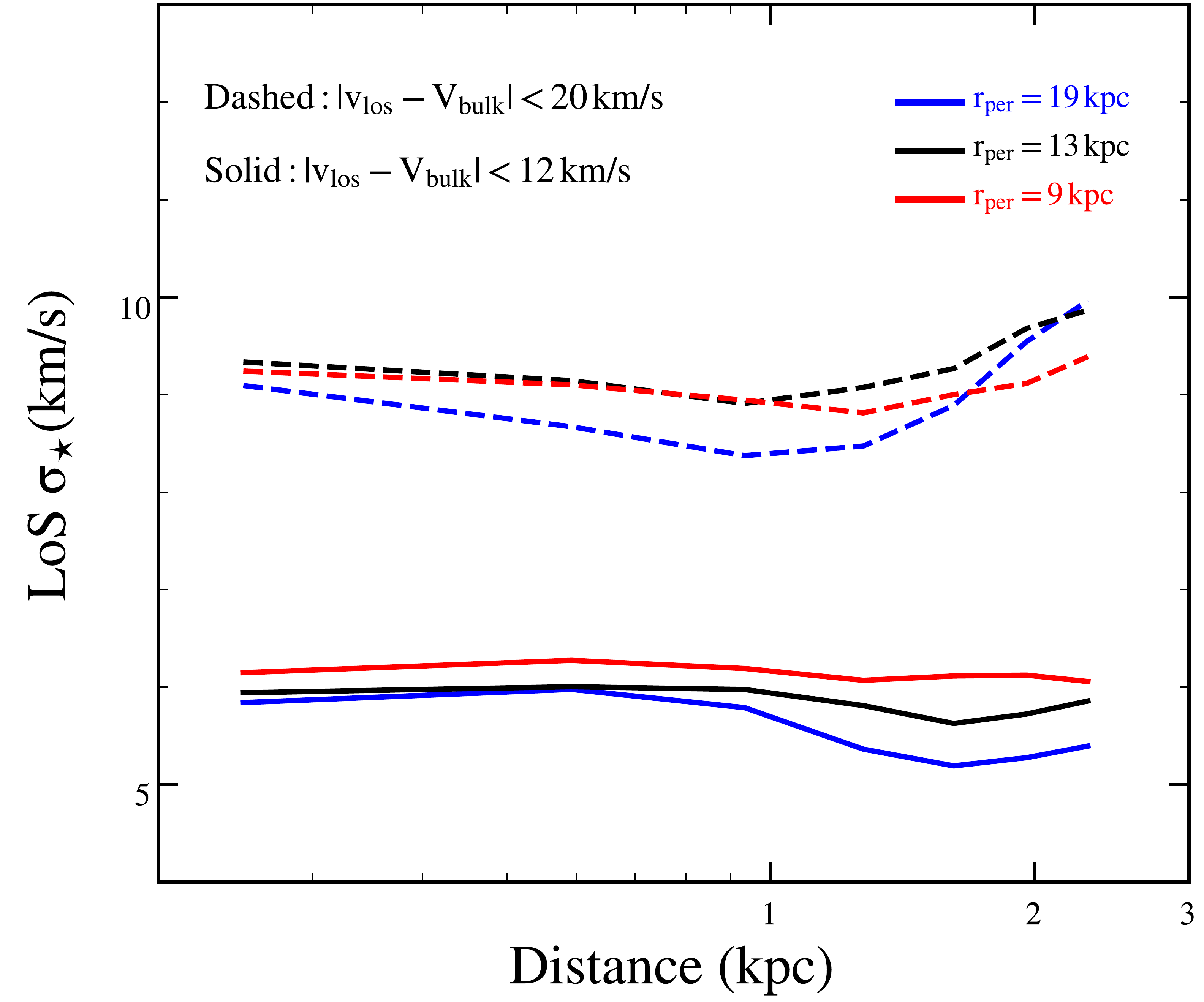}
    \caption{Left: LoS velocity dispersion of all stellar particles (solid) and only gravitationally bound particles (dashed) for the CDM (with $c_{\rm v}=10$) and SIDM3 models that pass the tight velocity cut $|v_{\rm los}-V_{\rm bulk}|<12~{\rm km/s}$. Right: LoS velocity dispersion for the CDM model ($c_{\rm v}=10$) with three different orbits, whose corresponding pericenters are $19~{\rm kpc}$, $13~{\rm kpc}$ and $9~{\rm kpc}$. Both loose (dashed) and tight (solid) velocity cuts are considered. }
    \label{fig:bound-vs-unbound}
\end{figure*}

Fig. \ref{fig:sigmaV} shows LoS velocity dispersion profiles for our CDM (left) and SIDM (right) simulations with the two velocity cuts, as well as the observed profile computed using the data compiled in \citet{torrealba2019}. It is clear that relaxing the cut leads to a significant increase in the velocity dispersion. The effect is more prominent for lower initial $c_{\rm v}$ or larger $\sigma/m$, as expected. This is also consistent with the results shown in Fig. \ref{fig:histogram}, i.e., the $20~{\rm km/s}$ cut reveals tidal debris structures in the stellar velocity distribution, and the galaxies with lower $c_{\rm v}$ or larger $\sigma/m$ have a relatively higher fraction of stellar particles on the tidal tail. Future observations  of the tidal debris of Antlia 2 would provide a means of testing those models.

We see that with the tight cut all the simulated galaxies reproduce consistently the observed LoS velocity distribution, although their dark matter densities inside a few $\rm kpc$ radii are different, as indicated in Fig. \ref{fig:density}. In fact, the resulting LoS $\sigma_\star$ profiles are remarkably similar; they are flat toward the center with little variation. It seems hard to use the measured LoS velocity dispersion of Antlia 2 to discriminate between the CDM and SIDM models in this work, because the simulated galaxies are tidally disturbed and their stellar particles do not provide a faithful tracer for the local mass distribution.

\label{sec:dynamicall-mass}

\begin{table}
    \centering
    \setlength\tabcolsep{5.5pt}
    \begin{tabular*}{\columnwidth}{lccc}
        \hline
         Name & $R_{1/2}\, ({\rm kpc}$) & ~$M_\star \, ({\rm M}_\odot)$ & ~$M(<R_{1/2})\, ({\rm M_\odot})$  \\
               \hline
         Antlia 2& $2.86\pm0.31$ & ~~~$(8.8\pm1.2){\rm e}5$ & $(5.4\pm2.1){\rm e}7$ \\ 
         \hline
         \hline
         CDM($c_{\rm v}$=13) &2.4&2.6e6  & 1.2e8\\
         CDM($c_{\rm v}$=10) &2.7&1.6e6&  6.0e7 \\ 
         CDM($c_{\rm v}$=8) &2.9&9.5e5& 3.2e7\\
         SIDM1&2.8& 1.4e6 & 5.4e7\\
         SIDM3 &3.0&1.0e6& 3.7e7\\
         SIDM5 & 3.2 &9.6e5 & 3.3e7\\
         \hline
    \end{tabular*}
    \centering
    \caption{Properties of the simulated satellites after $8~{\rm Gyr}$ of tidal evolution: the 2D projected half-mass radius of stars, the total stellar mass and the dynamical mass within $R_{1/2}$. For all simulated galaxies, the initial stellar mass is $3\times10^7~{\rm M_\odot}$ and the half-mass radius $R_{1/2}=2~{\rm kpc}$. The measurements of Antlia 2 \citep{torrealba2019} are also listed for comparison.} 
    \label{tab:sims}
\end{table}

\subsection{Stellar distributions and dynamical masses}

For each simulated galaxy, we compute its stellar half-mass radius and mass including all stars inside of $7~{\rm kpc}$ radius that pass the tight velocity cut and the results are summarized in Table \ref{tab:sims}. We see that $R_{1/2}$ expands relative to its initial value ($2~{\rm kpc}$), and the galaxies with shallower dark matter profiles, either lower $c_{\rm v}$ or higher $\sigma/m$, have more diffuse stellar distributions after tidal evolution \citep{errani2015,Dooley:2016ajo,Yang:2020iya}. It is also evident that the stars expand more significantly in SIDM than CDM, as the CDM halo retains its density cusp even if its concentration is as low as $c_{\rm v}=8$; see Fig.~\ref{fig:density}. All of our simulated galaxies have diffuse stellar distributions and their $R_{1/2}$ values are in good agreement with the observed half-light radius of Antlia 2, $R_{1/2}=2.86\pm0.31~{\rm kpc}$ \citep{torrealba2019} within the errors. Overall, our SIDM models match the observations slightly better than their CDM counterparts. It is important to note that for the halo mass we consider, our initial $R_{1/2}=r_{\rm p}=2~{\rm kpc}$ is already on the higher end of the stellar distribution predicted in simulations with realistic baryonic feedback \citep{lazar2020}. If the initial $R_{1/2}$ is smaller, SIDM would be more favored to reproduce Antlia 2's low surface brightness.

We compute the total stellar mass inside a radius of $7~{\rm kpc}$; see Table \ref{tab:sims}. The stellar mass decreases as the halo concentration decreases or the cross section increases, as expected. All of the galaxies have $M_\star$ comparable to Antlia 2's $M_\star=(8.8\pm1.2)\times10^{5}~{\rm M_\odot}$ \citep{torrealba2019} within a factor of two, and the models with lower $c_{\rm v}$ or larger $\sigma/m$ agree better with the data. If we choose a smaller stellar mass at the infall time, the resulting $M_\star$ would match the observed value closer. On the other hand, there could be additional uncertainties associated with the stellar mass measurements \citep{simon2019,applebaum2020}. Keeping those considerations in mind, we find the comparison to the stellar mass of Antlia 2 favors an under-dense halo. In addition, we have checked that the simulated stellar distributions can be well fitted by a Plummer profile and the relation $r_{1/2}=4/3R_{1/2}$ holds for our simulated galaxies to be better than $\sim7\%$ \citep[see also][]{Gonz_lez_Samaniego_2017,Campbell_2017}.

In Table \ref{tab:sims}, we also list the dynamical mass within $R_{1/2}$ for each simulated galaxy, computed from our simulations directly. There is a clear trend that $M_{\rm dyn}$ is reduced as $c_{\rm v}$ decreases or $\sigma/m$ increases. \cite{torrealba2019} estimated the dynamical mass of Antlia 2 enclosed within a sphere of radius $R_{1/2}\approx2.9~{\rm kpc}$ as $5.5\times10^7~{\rm M_\odot}$. It is based on the mass estimator $M(R_{1/2})=\mu \sigma^2_\star R_{1/2}$~\citep{walker2009}, where $\mu\equiv580~{\rm M_\odot~s^2/pc/km^{2}}$ and $\sigma_{\star}\approx 5.7~{\rm km/s}$ is the overall LoS $\sigma_\star$ of Antlia 2. The mass estimator assumes that the stars follow a Plummer profile, which is a good approximation for our galaxies. Taking the average values, $\sigma_\star\approx6~{\rm km/s}$ and $R_{1/2}\approx3~{\rm kpc}$, we apply the estimator to our galaxies and find $M({<R_{1/2}})\sim6\times 10^7~{\rm M_\odot}$, in good agreement with the estimated one for Antlia 2. 

However, compared to the results directly from our simulations, the mass estimator underestimates the dynamical mass for the CDM halo with $c_{\rm v}=13$ by a factor of $2$, while overestimates it for SIDM5 by the same factor. As shown in Fig.~\ref{fig:sigmaV}, all the galaxies have approximately the same LoS $\sigma_\star$ profile after the tight velocity cut. For the CDM halo ($c_{\rm v}=13$), our cut removes stellar particles on the high velocity tail, resulting in relatively low $\sigma_\star$; see Fig. \ref{fig:histogram} (left). If we were taking $\sigma_\star\approx10~{\rm km/s}$ from the loose cut, the agreement would be much better for the CDM halo. On the other hand, SIDM5 has the shallowest potential and its stellar orbits are disturbed most significantly, leading to LoS $\sigma_\star$ much higher than that expected without the tidal effects. The discrepancy is not surprising, because the mass estimator is based on Jeans modeling, which assumes a system in dynamical equilibrium, but SIDM5 is least equilibrated among the six simulated galaxies.

\subsection{Bound vs. unbound stellar particles}

\label{sec:bound}
All dark matter and stellar particles are gravitationally bound to the simulated satellites at $t=0$ by construction, while subsequent tidal interactions with the host potential will unbind significant amounts of mass. Loosely bound or unbound particles are more frequently stripped at outer radii of the galaxies. In computing the LoS velocity dispersion shown previously, we did not exclude the unbound stellar particles, because from the observational perspective it is not possible to single out unbound stars. It is interesting to check how the LoS velocity dispersion profile change if we only include bound stellar particles.

Fig. \ref{fig:bound-vs-unbound} (left) shows LoS $\sigma_\star$ profiles for the CDM model with $c_{\rm v}=10$ and the SIDM3 model around $8~{\rm Gyr}$ when the unbound particles are included or excluded, where we adopt the tight velocity cut, $|v_{\rm los}-V_{\rm bulk}|\leq 12~{\rm km/s}$. Interestingly, ignoring the contribution from the unbound particles only slightly decreases the LoS velocity dispersion in the CDM model, while in the SIDM3 model it reduces by more than a factor of two. This is because the latter has a shallower gravitational potential and most of stellar particles become unbound after tidal evolution. Thus, we expect to see more tidal streams in the SIDM models.

\subsection{Impact of different pericenters}
\label{sec:pericenter}

After sampling observed errors of the Gaia DR-2 proper motions, \cite{chakrabarti2019} derived an orbital distribution for the Antlia 2 galaxy. The resulting pericenter is in the range of $8\textup{--}30~{\rm kpc}$, with a median around $15~{\rm kpc}$  for a typical mass for the Milky Way halo. The orbit that we have chosen has a pericenter of $13~{\rm kpc}$. It is interesting to investigate how our results change if we choose other orbits from this orbital distribution.

We focus on the impacts on the LoS velocity dispersion and perform two additional simulations for the CDM model with $c_{\rm v}=10$. The corresponding orbits have pericenters of $9~{\rm kpc}$ and $19~{\rm kpc}$, which are within $1\sigma$ of the orbital distribution derived from Gaia DR-2 data \citep{chakrabarti2019}. Fig. \ref{fig:bound-vs-unbound} (right) shows the LoS stellar velocity dispersion for the three orbits for the CDM model with $c_{\rm v}=10$. The LoS $\sigma_\star$ profiles change very mildly with the pericenter for the CDM model. It would be interesting to explore the impact of different pericenters on SIDM haloes.

\section{Conclusions}

\label{sec:conclusion}

We have performed N-body simulations to model the recently discovered Antlia 2 galaxy. The simulation suite consists of three CDM haloes with halo concentrations $c_{\rm v}=13,\, 10,$ and $8$, and three SIDM haloes with $c_{\rm v}=10$ and $\sigma/m=1,\, 3$ and $5~{\rm cm^2/g}$. The simulated dwarf galaxies start with an NFW and Plummer profiles for their dark matter and stellar components, respectively, with well-motivated model parameters. 
Our main findings are: 

\begin{itemize}

 \item After $8~{\rm Gyr}$ of tidal evolution, the CDM haloes remain cuspy in the inner regions even if the initial concentration is as low as $c_{\rm v}=8$, while the SIDM haloes develop dark matter density cores and their sizes increase with the cross section.

 \item The Milky Way's tidal field strips most of the dark matter and star particles, resulting in significant tidal debris for the simulated galaxies. The stripping increases with decreasing $c_{\rm v}$ or increasing $\sigma/m$. The stellar velocity distribution exhibits a feature of tidal ``arms" (Fig. \ref{fig:histogram}). If this is verified in the future observations of Antlia 2, it would serve as an important test of this scenario. In general, SIDM haloes predict more tidal debris than their CDM counterparts.  
 
 \item The simulated LoS velocity dispersion agrees with the observed one after imposing the tight cut, $|v_{\rm los}-V_{\rm bulk}|\leq 12~{\rm km}$. All our simulated galaxies have similar stellar dispersion profiles, regardless the initial concentration and the self-scattering dark matter cross section. 
    
 \item The half-mass radius of the simulated galaxies expands after tidal evolution. The haloes with less dense central densities, either due to low $c_{\rm v}$ or high $\sigma/m$, allow the stellar distribution to become more diffuse, and the stellar mass decreases accordingly; see Table \ref{tab:sims}. Our results indicate Antlia 2 has an under-dense progenitor.
    
 \item Similar to the stellar mass, the dynamical mass within the half-mass radius, dominated by dark matter, also correlates with the initial halo concentration and the SIDM cross section. For a system with a shallow density profile, its stellar orbits can be significantly disturbed in the tidal field and the mass estimator based on the Jeans modeling can significantly overestimate its dynamical mass.
    
 \item The simulated CDM galaxies contain more gravitationally bound stellar particles than their SIDM counterparts, as the former has a deeper potential. The velocity dispersion profile is insensitive to the pericenter for the CDM model that we have checked. 
\end{itemize}

Considering the above observations, we find that Antlia 2 could be realized in both CDM and SIDM with well-motivated model parameters. In our simulations, we chose the initial stellar and halo masses using the mass-metallicity and abundance-matching relations, respectively, and adopted the stellar scale radius consistent with predictions from hydrodynamical simulations with realistic feedback prescriptions. We find that the initial halo concentration in CDM should be lower than the cosmological mean to be consistent with the observations of Antlia 2. In SIDM, core formation enhances the tidal mass loss for both dark matter and stars; it also enhances the expansion of the stellar distribution. SIDM, therefore, allows a higher halo concentration and a denser stellar distribution compared to CDM for models that match observations of Antlia 2. In this work, we used controlled simulations with a static host potential to study the key ingredients to realize Antlia 2-like galaxies. It would be of great interest to further explore this topic with cosmological hydrodynamical simulations.

\section*{Acknowledgements}
We thank Volker Springel for providing {\sc Arepo} for this work, Josh Simon and Ting Li for useful discussions. SC acknowledges support from
NASA ATP NNX17AK90G, NSF AAG grant 1517488, and from Research Corporation for Scientific Advancement's Time Domain Astrophysics Scialog. HBY acknowledges support from U.S. Department of
Energy under Grant No. de-sc0008541, and NASA grant 80NSSC20K0566. MBK acknowledges support from NSF CAREER award AST-1752913, NSF grant AST-1910346, NASA grant NNX17AG29G, and HST-AR-15006, HST-AR-15809, HST-GO-15658, HST-GO-15901, and HST-GO-15902 from the Space Telescope Science Institute, which is operated by AURA, Inc., under NASA contract NAS5-26555. MV acknowledges
support through an MIT RSC award, a Kavli Research
Investment Fund, NASA ATP grant NNX17AG29G,
NSF grants AST-1814053 and AST-1814259. JZ acknowledges support by a Grant of Excellence from the
Icelandic Research Fund (grant number 173929). 

\bibliographystyle{mnras}
\bibliography{bib}


\bsp	
\label{lastpage}
\end{document}